\documentclass[11pt]{article}
\usepackage{moriond,epsfig}

\def\be{\begin{equation}}
\def\ee{\end{equation}}
\def\bea{\begin{eqnarray}}
\def\eea{\end{eqnarray}}

\def\numunue{\nu_\mu\rightarrow\nu_e}
\def\numunutau{\nu_\mu\rightarrow\nu_\tau}

\def\nutau{\nu_\tau}

\def\numu{\nu_\mu}

\def\tautoe{\tau \rightarrow e}
\def\tautomu{\tau \rightarrow \mu}
\def\tautoh{\tau \rightarrow h}

\begin{document}

\vspace*{4cm}
\title{OPERA first events from the CNGS neutrino beam}

\author{ J.Marteau for the OPERA collaboration }

\address{Institut de Physique Nucléaire de Lyon - UCBL-IN2P3, 4 rue E.Fermi,\\
69622 Villeurbanne, France}

\maketitle\abstracts{
The aim of the OPERA experiment is to search for the appearance of the tau neutrino 
in the quasi pure muon neutrino beam produced at CERN (CNGS). The detector, installed 
in the Gran Sasso underground laboratory 730~km away from CERN, consists of a lead/emulsion target complemented 
with electronic detectors. A report is given on the detector status 
(construction, data taking and analysis) and on the first successful 2006 neutrino runs.}

\section*{Introduction}
In the last decades solar and atmospheric neutrino experiments observed deficits in the measured fluxes which are all
well reproduced in a neutrino oscillations model, implying non vanishing, not degenerate neutrino masses and neutrino mixing.
Within such hypothesis weak interactions eigenstates differ from the mass eigenstates. The mixing can be parametrized in an unitary
matrix whose parameters (3 angles and 1 or 3 phases depending on the Dirac or Majorana nature of neutrinos) associated to the square
masses differences $\Delta m^2$ drive the amplitude of the disappearance ($P(\nu_\alpha \rightarrow \nu_\beta)$) or survival 
($P(\nu_\alpha \rightarrow \nu_\alpha)$) probabilities. The major experimental results for solar neutrinos come from radiochemic experiments\cite{chlore,gno,sage} or large \v{C}erenkov detectors\cite{sk,kam,imb,sno}. For atmospheric neutrinos they come from \v{C}erenkov and calorimeters~\cite{soudan2,macro}. These results were confirmed with reactor~\cite{chooz,kamland} and long baseline experiments~\cite{k2k,minos}. All these experiments are however of ``disappearance" type, they compare the measured flux at a far distance with either one at a close position (in the case of long baseline experiments) or with the predicted one (in the reactor, ``solar" and ``atmospheric" experiments). 

The OPERA experiment~\cite{opera} has been designed to perform an unique \underline{appearance} observation of the oscillation products to confirm (or infirm) the neutrino oscillation hypothesis in the atmospheric sector through the $\numunutau$ channel and also to set limits on the $\theta_{13}$ angle through the $\numunue$ channel. This article reports the first observed neutrino events from the CNGS (CERN to Gran Sasso) beam by the OPERA experiment.   

\section{The OPERA experiment}
\subsection{The CNGS programme}\label{subsec:cngs}
The CNGS~\cite{cngs} programme of neutrino beam from CERN to Gran Sasso has been approved in 1999. OPERA was approved as ``CNGS1 experiment'' in 2001. From CERN to Gran Sasso the neutrinos time of flight is 2.44~ms and their average direction makes a 3$^\circ$ angle w.r.t. the horizontal due to the earth curvature. 
The main features of the beam have been presented in details in this conference~\cite{edda}. The beam has been optimized to maximize the number of $\tau$ events in the detector (convolution of the neutrino flux, the disappearance probability $P(\numunutau)$ and the detection efficiency). The neutrino average energy is 17~GeV. The $\bar{\nu_\mu}$ contamination is $\sim 4\%$, the $\nu_e$ ($\bar{\nu_e}$) is $< 1\%$ and the number of $\nu_\tau$ is negligible. The expected beam intensity is $4.5 \cdot 10^{19}$ p.o.t./year.

\subsection{The detection technique}\label{subsec:ecc}
The challenge of the experiment is to measure the appearance of $\nutau$ from $\numu$ oscillations through CC $\tau$ interactions. The events induced by the short-lived $\tau$ have a characteristic topology (with a ``kink'' due to the presence of undetected neutrinos in the $\tau$ decay) but extends over $\sim mm^3$ typical volumes. 

\paragraph{ECC technique} The detector should therefore match a large mass for statistics, a high spatial resolution and high rejection power to limit background contamination. These requirements are satisfied using the proven ECC (Emulsion Cloud Chamber) technique which already worked successfully in the DONUT experiment~\cite{donut}. The passive target consists of lead plates. Particles are tracked in nuclear emulsions films with a sub-micrometric intrinsic resolution. 57 emulsions films are assembled and interspaced with 56 lead plates 1~mm wide in a detector basic cell called ``brick''. A brick is a 12.7 $\times$ 10.2 cm$^2$ object with a thickness along the beam direction of 7.5 cm (about 10 radiation lengths). Its weight is about 8.3~kg.
Bricks are assembled in 31 walls ($52 \times 64$ bricks) separated by electronic detectors planes to trigger the event and identify the brick with the interaction vertex. 

The readout sequence of OPERA events is quasi-online. Once identified the brick is extracted from the detector, emulsions are developed and scanned by automatic microscopes. Scanning performs detailed tracking, vertex location, particle identification, momentum measurement through Multiple Coulomb Scattering, decay kink search. The data are complemented with the momentum, energy and charge measurements done by the electronic detectors.   

\paragraph{Expected performances} At the expected nominal beam intensity and for five years data taking a total of 31000 charged and neutral current interactions is expected in the nominal mass target of OPERA. 
Among these 95 (214) CC $\nutau$ interactions are expected for oscillation parameter values $\Delta m^2_{23}$=2~(3)~$\times$~10$^{-3}$~eV$^2$ and sin$^2$2$\theta_{23}$=1. The overall detection efficiencies have been estimated in Monte Carlo simulations upgraded by dedicated tests (vertex searches, $e/\pi$ and $\pi/\mu$ separation, large angle muon scattering...). The nominal expected number of $\tau$ events ranges from 11 to 16 events with the same parameters sets. The physics channel considered so far are $\tautoe$, $\tautomu$ and $\tautoh$ (1 or 3 prongs). The background is expected to be $< 1$. 
 In the sub-dominant $\numunue$ channel OPERA should, in the same run conditions, set a limit of $\sin^2 2\theta_{13}<0.06$ (90\%~C.L.) \cite{t13bib} assuming $\Delta m^2 = 2.5 \cdot 10^{-3}$ eV$^2$.   
 
\subsection{The OPERA detector}\label{subsec:detector}
\begin{figure}[!ht]
\center{\psfig{figure=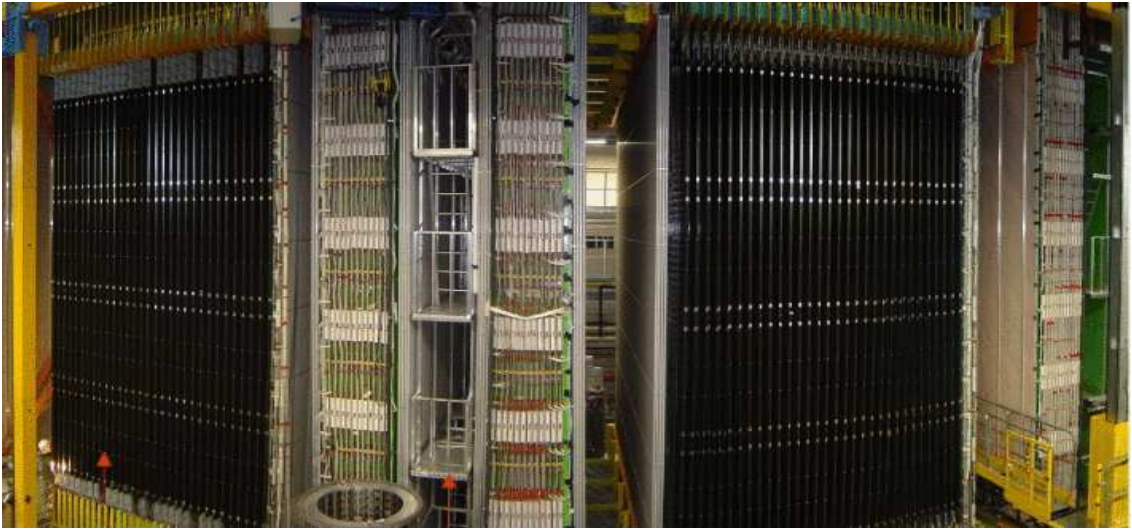,height=7cm,width=15cm}}
\caption{Side view of the OPERA detector. The two target modules, 31 bricks walls spaced by black 
covered scintillator biplanes, are separated by the first spectrometer, 6 stations of drift tubes upstream, between
and downstream the two arms of a dipolar magnet instrumented with RPC.  
\label{fig:opera}}
\rule{5cm}{0.2mm}\hfill\rule{5cm}{0.2mm}
\end{figure}

The OPERA detector is divided into two Super-Modules consisting of a target section followed by a muon spectrometer (see Picture~\ref{fig:opera}). The target section is made of 31 brick walls each one being followed by a highly segmented scintillator tracker plane. A large VETO plane is placed in front of the detector to further discriminate beam events from horizontal cosmics. The construction of the experiment started in Spring 2003. The two instrumented magnets were completed in May 2004 and beginning of 2005 respectively. In Spring 2006 all scintillator planes were installed.  

\paragraph{The electronic detectors}
The target tracker covers a total area of 7000~m$^2$ and is built of 32000 scintillator strips, each 7~m long and of 25~mm$ \times $15~mm cross section. Along the strip, a wavelength shifting fiber of 1~mm diameter transmits the light signals to both ends, read out by 992 multi-anode (64 channels) PMT's from Hamamatsu. A 32 channel front-end electronics ASIC~\cite{asic} has been developed which allows individual gain corrections (with a dynamic range of $1 \div 4$) and auto-triggered readout sequence in a standard dual shaper scheme (fast and slow) and track~\&~hold logic.

The muon spectrometer consists of a large $8 \times 8$~m$^2$ dipolar magnet delivering a magnetic field of 1.55~T and instrumented with RPC's and drift tubes. Each magnet arm consists of twelve 5~cm thick iron slabs, alternating with RPC planes. This sandwich structure allows the tracking in the magnetic field to identify the muons and to determine their momentum. In addition the precision tracker \cite{specbib} measures the muon track coordinates in the horizontal plane. It is made of 8~m long drift tubes with an outer diameter of 38~mm. The charge misidentification is expected to be 0.1~\% - 0.3~\% in the relevant momentum range which is efficient enough to minimize the background originating from the charmed particles produced in $\nu_{\mu}$ interactions. With the muon spectrometer a momentum resolution of $\Delta p / p \le 0.25$ for all muon momenta $p$ up to a maximum of $p=25$~GeV/c can be achieved. 

\paragraph{"Bricks" production}
In total $\sim$~200000 bricks should be nominally produced and installed in OPERA. The production is performed by a dedicated apparatus called Brick Assembly Machine (BAM) installed in Gran Sasso. It is a chain of different stations (for piling, pressing, wrapping etc) using robots operating in dark rooms with controlled environment. The production is around $\sim$~2 bricks per minute. 
Once bricks are produced they are placed in dedicated mechanical structures called drums (9 rows of 26 bricks) and inserted from there to the detector by the Brick Manipulating System (BMS). The BMS has two equivalent structures (one per side) consisting of a brick storage place (carousel) where the drums are exchanged and a moving robot along the side of the experiment. The mobile part of the BMS can reach the desired row and plane with a sub-millimetric accuracy. The robot has a mobile bridge on which the bricks are pushed inside the detector by a pushing arm to their desired position. The brick extraction is performed by a vacuum sucker located on the front of a small vehicle.    

\section{Data taking and analysis}
\paragraph{DAQ and on-line analysis}
The OPERA DAQ system is based on a so-called ``smart'' sensor concept on an Ethernet network. The principle is to implement a local micro-processor as close as possible to the front-end electronics and to access it for configuration and/or data transmission through Ethernet. The core of this architecture is a small processor board which hosts a sequencer (FPGA of the Altera cyclone family), a micro-processor (32 bits RISC Etrax100lx processor from Axis) and an intermediate FIFO. The FPGA manages the full readout sequence, the data timestamping with a 10~ns accuracy, the data formatting and pre-processing (pedestals and zero suppression, local histogramming) and the data transfer to the intermediate buffer. The processor runs ``sensor'' applications communicating with ``daq'' servers and developed within the CORBA framework implemented in C++.  

Each sensor is plugged to a sub-detector specific motherboard and is seen as a standard Ethernet node over a large network. In total 1153 sensors are connected for a total of 100000 readout channels. The synchronization of each individual clock is performed through a specific bi-directional bus starting from a GPS PCI board developed on purpose. A common 20~MHz clock embedding specific signals is send to the sensors with a measurement of time propagation delays for off-line correction. 

Data analysis and reconstruction is performed continuously on-line within the Opera software framework based on ROOT.
 
\paragraph{Off-line emulsion scanning}
After development emulsions are scanned by automatic microscopes whose nominal speed is higher than $\sim 20$ cm$^2$/h per emulsion layer ($44 ~\mu$m thick). There are two different approaches developed by the OPERA collaboration, in Europe (ESS~\cite{ESS}, based on software image reconstruction) and in Japan (S-UTS~\cite{SUTS} based on hard-coded algorithms). Next picture displays an example of both systems. 
The scanning sequence proceeds with the division of the emulsion thickness into $\sim$~16 tomographic images by focal plane adjustment, images digitization and track finding algorithms. Track grains are identified and separated from ``fog'' grains and associated into ``micro-tracks''. Examples of reconstructed tracks during 2006 runs will be given below.
\begin{figure}[!hb]
\center{\psfig{figure=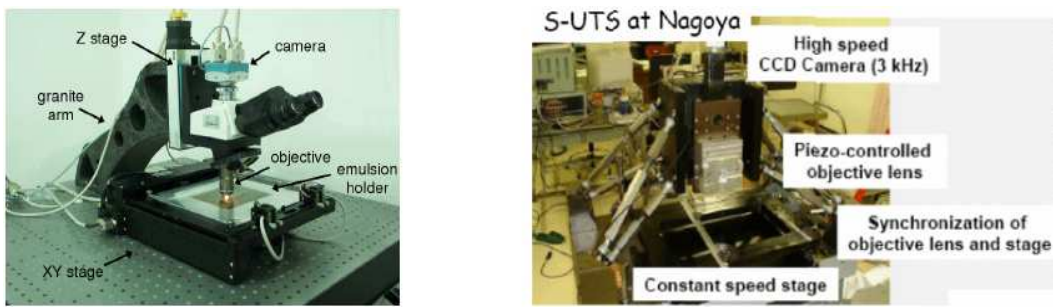,height=4cm,width=14cm}}
\caption{Pictures of one of the ESS microscopes (left) and of the
  S-UTS (right).  
\label{fig:scanning}}
\rule{5cm}{0.2mm}\hfill\rule{5cm}{0.2mm}
\end{figure}

\section{First CNGS neutrino events}
\begin{figure}[!ht]
\center{\psfig{figure=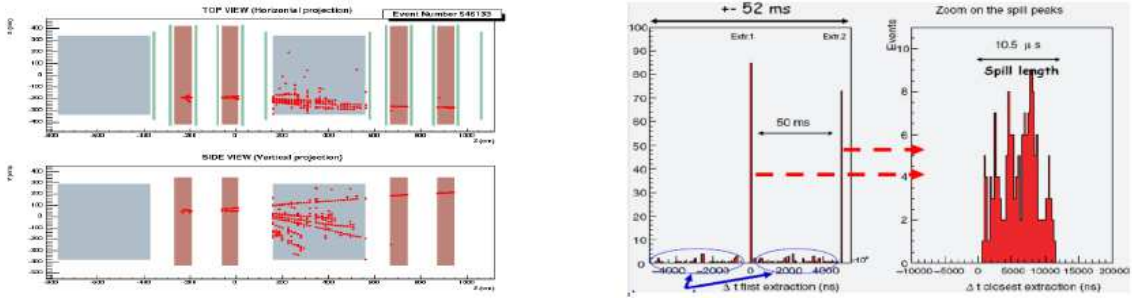,height=4cm,width=15cm}}
\caption{Left: event display of a neutrino interaction in the spectrometer. Right : timing distribution of beam related events. The 2 CNGS fast extractions, separated by 50~ms are clearly reconstructed. The right plot focuses on one of the two extractions. The typical width reconstructed ($\sim10\mu$s) is coherent with the expectations. 
\label{fig:time}}
\rule{5cm}{0.2mm}\hfill\rule{5cm}{0.2mm}
\end{figure}

\paragraph{Beam structure reconstruction}
During the first CNGS run in August 2006 319 neutrino events were collected with an estimated systematic error of 5\% (see Fig.\ref{fig:time} left for a typical display). The events were selected by a comparison of their absolute timestamps w.r.t. the beam time information in quite large coincidence window (1ms). The beam spill time structure reconstructed in OPERA is displayed in Fig.\ref{fig:time} (right).   

\begin{figure}[!ht]
\center{\psfig{figure=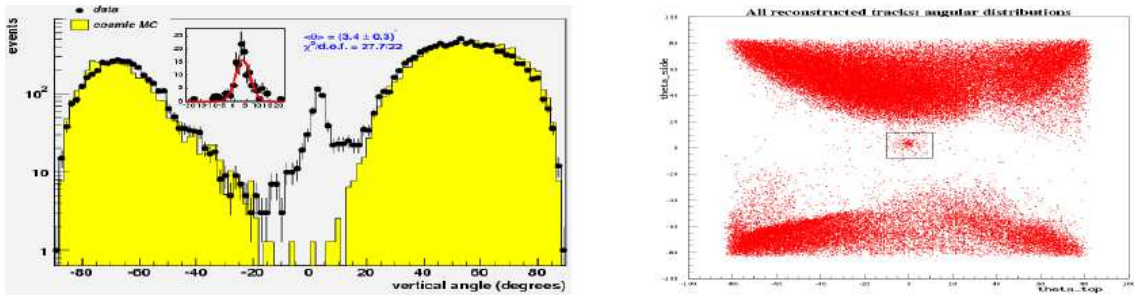,height=4cm,width=15cm}}
\caption{Left: angular distribution w.r.t. horizontal direction of events reconstructed with the spectrometer RPC (black points) compared to MC expectations for cosmics only (yellow histogram). Right: scatter plot of side and top projection angles wrt horizontal 
direction reconstructed with the target tracker. Beam events appear in the central zone of those distributions.  
\label{fig:angle}}
\rule{5cm}{0.2mm}\hfill\rule{5cm}{0.2mm}
\end{figure}

\paragraph{Cosmics vs neutrino events}
Beam events have an average direction close to the horizontal one ($3.3^\circ$ angle) whereas cosmics have large angles distributions. This is shown in the distributions of Fig.~\ref{fig:angle}. A gaussian fit to the central distribution leads to a mean angle of 3.4$\pm$0.3$^\circ$ in agreement with the expected value.

\begin{figure}[!ht]
\center{\psfig{figure=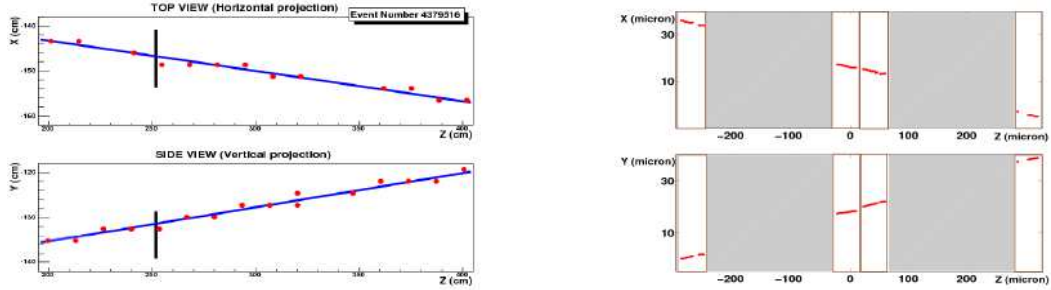,height=4cm,width=14cm}}
\caption{Left: display of one event with the muon passing through the emulsion sheet detector plane. Right: display of the corresponding 4 micro-tracks reconstructed in the emulsion.  
\label{fig:cstt}}
\rule{5cm}{0.2mm}\hfill\rule{5cm}{0.2mm}
\end{figure}

\paragraph{Emulsions matching} Some clean tracks have been followed in emulsion sheets inside the detector. A typical display of the hits reconstructed in the emulsions and in the TT is given in Fig.~\ref{fig:cstt}.  
\pagebreak
\section*{Conclusions and perspectives}
OPERA performed the first detection of neutrino events from the long baseline CERN CNGS beam in the underground Gran Sasso laboratory. 319 neutrino-induced events were collected for an integrated intensity of 7.6 $\times$ 10$^{17}$ p.o.t. in agreement with the expectations. The reconstructed zenith-angle distributions and the time structure of the events demonstrate the capability of the electronic detectors, build up during the last three years, to reach the experiment goals. The association of tracks between electronic detectors and emulsion sheets has been also successfully performed. The collaboration is facing the last large effort of brick production and insertion and is preparing next neutrinos runs in fall 2007 for physics commissioning. 


\section*{Acknowledgments}
I would like to acknowledge the cooperation of all the members of the OPERA
Collaboration and thank the organizers of the conference for the invitation.

\end{document}